\documentclass[aps,pra,twocolumn,amsmath,superscriptaddress]{revtex4-2}

\newcommand{\bea}{\begin{eqnarray}}
\newcommand{\eea}{\end{eqnarray}}
\newcommand{\beq}{\begin{equation}}
\newcommand{\eeq}{\end{equation}}

\newcommand{\ep}{\epsilon}

\newcommand{\Si}{\Sigma}

\usepackage[urlcolor=blue,colorlinks=true,citecolor=blue,linkcolor=blue,pdfstartview={FitH},bookmarks=false]{hyperref}
\usepackage{graphicx}
\usepackage{longtable}
\usepackage{epsfig}
\usepackage{dcolumn}
\usepackage{bm}
\usepackage{amssymb}
\usepackage{multirow}
\usepackage{times,color}
\usepackage{hyperref}
\usepackage{graphics}
\usepackage{bbm}
\usepackage{amsfonts}
\usepackage{subfigure}
\usepackage{float}
\usepackage{CJK}
\begin{document}

\title{Quantum Many-Body Scars in Spin-1 Kitaev Chains}

\author{Wen-Long You}%\email{wlyou@nuaa.edu.cn}
\affiliation{College of Science, Nanjing University of Aeronautics and Astronautics, Nanjing, 211106, China}
\affiliation{Key Laboratory of Aerospace Information Materials and Physics (NUAA), MIIT, Nanjing 211106, China}

\author{Zhuan Zhao}
\affiliation{College of Science, Nanjing University of Aeronautics and Astronautics, Nanjing, 211106, China}
\affiliation{Key Laboratory of Aerospace Information Materials and Physics (NUAA), MIIT, Nanjing 211106, China}

\author{Jie Ren}%\email{jren@cslg.edu.cn}
\affiliation{Department of Physics, Changshu Institute of Technology, Changshu 215500, China}

\author{Gaoyong Sun}
\affiliation{College of Science, Nanjing University of Aeronautics and Astronautics, Nanjing, 211106, China}
\affiliation{Key Laboratory of Aerospace Information Materials and Physics (NUAA), MIIT, Nanjing 211106, China}

\author{Liangsheng Li}
\affiliation{Science and Technology on Electromagnetic Scattering Laboratory, Beijing 100854, China}

\author{Andrzej M. Ole\'s$\,$}
\email{a.m.oles@fkf.mpi.de}
\affiliation{\mbox{Max Planck Institute for Solid State Research,
             Heisenbergstrasse 1, D-70569 Stuttgart, Germany} }
\affiliation{\mbox{Institute of Theoretical Physics, Jagiellonian University,
             Prof. Stanis\l{}awa \L{}ojasiewicza 11, PL-30348 Krak\'ow, Poland}}

\begin{abstract}
To provide a physical example of quantum scars, we study the many-body
scars in the spin-1 Kitaev chain where the so-called PXP Hamiltonian is
exactly embedded in the spectra. Regarding the conserved quantities, the
Hilbert space is fragmented into disconnected subspaces and we explore
the associated constrained dynamics. The continuous revivals of the
fidelity and the entanglement entropy when the initial state is prepared
in $\vert\mathbb{Z}_k\rangle$ ($k=2,3$) state illustrate the essential
physics of the PXP model. We study the quantum phase transitions in the
one-dimensional spin-1 Kitaev-Heisenberg model using the density-matrix
renormalization group and Lanczos exact diagonalization methods, and
determine the phase diagram.
We parametrize the two terms in the Hamiltonian by the angle $\phi$,
where the Kitaev term is $K\equiv\sin(\phi)$ and competes with the
Heisenberg $J\equiv\cos(\phi)$ term. One finds a rich ground state
phase diagram as a function of the angle $\phi$.
Depending on the ratio $K/J\equiv\tan(\phi)$, the system either breaks
the symmetry to one of distinct symmetry broken phases, or preserves the
symmetry in a quantum spin liquid phase with frustrated interactions.
We find that the scarred state is stable for perturbations which obey
$\mathbb{Z}_2$-symmetry,
while it becomes unstable against Heisenberg-type perturbations.\\
\textit{Accepted for publication in Physical Review Research}
%\keywords{quantum phase transitions \and string order \and Kitaev-Heisenberg model }

\end{abstract}

\date{20 January 2022}

\maketitle

\section{Introduction}
\label{intro}

With the rapid development of experimental techniques, specifically
with the advancement of stable and precisely adjustable ultrashort
pulse laser technologies, and the improvement of detection methods,
the real-time information with atomic scale and sub-picosecond
resolution can be obtained. It provides new opportunities for
perceiving the phenomena in related many-body systems.
Meanwhile, quantum simulators with high isolation, long relaxation
time and experimental controllability, such as ultra-cold atoms,
ion traps, nitrogen vacancy centers in diamonds, and superconducting
circuits, have opened up a new direction in the studies of strongly
correlated electron systems in condensed matter physics. Studies of
abstract quantum complex systems is another route to provide a more
complete description of quantum phase transitions and many-body
dynamics in isolated quantum systems.

Particularly in the recent years, a tremendous progress was made
in describing the non-equilibrium dynamics of isolated quantum
systems. Thermalization and localization are distinct fates of the
steady quantum states. The former will lose information of the
initial state after undergoing a long-time evolution, while the latter
retains information of the initial state as much as possible.
In general, the system will tend to thermalize, as it follows
the eigenstate thermalization hypothesis (ETH), which is generally
regarded as the cornerstone of contemporary quantum statistics. A system
that satisfies ETH is considered to be ergodic. Numerical calculations
show that most interacting systems are indeed strongly
ergodic~\cite{Abanin19,Rigol,Rigol08,Polkovnikov11}. Thermal excitations
enrich the properties of spin chains and are, for instance, responsible
for the dimerization in spin-orbital systems \cite{Sir08}.

However, currently many quantum systems were found to disobey the ETH.
A typical exception is the integrable system undergoing Anderson
localization \cite{Anderson58}, in which there are infinite local
conserved quantities that can inhibit the relaxation between eigenstates
and thus retain the coherence in the quantum system. However even a weak
perturbation will suffice to break the integrability of the system, and
then the ETH would be restored~\cite{Brenes20}.

Another mechanism to avoid thermalization is many-body localization---it
has been intensely investigated in the past decade
\cite{Basko06,Pal10,Znidaric08,Bar16}.
It was demonstrated that using one-dimensional (1D) so-called PXP models
for spin-1 can lead to weak ergodicity breaking \cite{Muk21}. The
emergent integrability of these \textit{a priori} non-integrable systems
prevents them from thermalization \cite{Imbrie16}. The Anderson
localization generally concerns the ground state, while the many-body
localization usually occurs at finite temperature. Recently, the phase
transitions between the thermalized and many-body localized phases,
highlighting many-body localized mobility edge, were observed in
19-qubit superconducting systems~\cite{Guo21}.

Nowadays exact quantum many-body scars (QMBSs) and their stability in
constrained quantum chains attract attention~\cite{Sur21}. Quantum scars
are nonthermal eigenstates characterized by low entanglement entropy,
initially detected in systems subject to nearest neighbor Rydberg
blockade, the PXP model \cite{Fen04,Les12,Tur182,Turner18}.
This model has been introduced as a simplified description of the
quantum simulator on 51 Rydberg atoms \cite{Bernien17}. It has been
shown that the quantum quench dynamics starting from an
antiferromagnetic (AFM) state will return to the certain initial state
repeatedly after reaching equilibrium. The continuous
oscillations of physical observables such as domain-wall density show
that the system exhibits a new type of non-thermalization behavior and
the peculiar behavior was dubbed as quantum many-body scars (QMBSs). If
the initial state is prepared in other generic states, the atomic
system will thermalize as expected. In addition to the Rydberg atomic
system, QMBSs were also observed in 1D dipolar Bose gas~\cite{Kao21}.

The ideas of quantum scar state were borrowed from the classic chaotic
scar phenomenon, which is characterized by a high probability density
distribution near a specific classical periodic orbit. Quantum scars
were originally proposed by Heller in 1984 based on single-particle
picture \cite{Heller1984}. QMBSs are abnormal condensations of quantum
many-body wave functions on specific eigenstates, and these eigenstates
constitute quantum scars. On one hand, because QMBSs are different from
the conventional paradigm of thermalization, Anderson localization and
many-body localization, the nature of this novel quantum phenomenon has
stimulated the interest of many physicists. On the other hand, QMBSs may
represent a new way to realize coherent quantum mechanics, which can
protect the quantum dense coding processing for a long time. Obviously,
searching for non-thermal dynamics in interacting non-integrable
quantum systems has not only fundamental theoretical significance
in condensed matter theory, but also has great practical significance
in quantum information processing.

In a nutshell, the scarred model is characterized by a subspace which
is decoupled from the rest of the energy spectrum and cannot be simply
attributed to a symmetry of the system. Taking the spin-$S$ system on a
chain with length $N$ as an example, the dimension of the full Hilbert
space scales exponentially as $(2S+1)^N$, and the dimension of the scar
subspace scales algebraically with the system size $N$ as ${\cal O}(N)$.
These zero-measure scar eigenstates allow the quantum system to maintain
long-time coherent dynamics. The underlying mechanism for the 51 Rydberg
atoms finding the return journey of a specific initial state in the
Hilbert space of more than $4\times 10^{10}$ dimensions is fuzzy.

Ergodicity breaking in such systems can often be attributed to the
presence of symmetries (hidden, emergent, or explicit) that preclude
the establishment of a global equilibrium state~\cite{Pakrouski20,JRen21}.
The existence of quantum scars may be revelent to emergent integrability
of Hamiltonian \cite{Khemani19}, the algebraic structure of dynamic
symmetry \cite{Buca19}, while some researches rule out the integrability
\cite{Choi19}.

A flurry of subsequent theoretical works have addressed
the puzzle posed by the experiment: the nature and origin of this new
regime of ergodicity breaking, intermediate between thermalization and
strong ergodicity breaking. Notably, an extensive research on weak
ergodicity breaking has been performed in various quantum many-body
systems, such as extended Ising model \cite{Iadecola20}, integer spin XY
model \cite{Chattopadhyay20}, Affleck-Kennedy-Lieb-Tasaki (AKLT) model
\cite{Moudgalya18,Mark20a} and the Hubbard model
\cite{Mou20,Mark20b,Desaules21}.

Kitaev models have attracted much attention currently as they
provide deeper understanding of the spin-orbital physics in transition
metal oxides~\cite{Nus15}. The search for their phase diagrams is
motivated by a topological quantum spin liquid (QSL) ground state and
Majorana excitations emerging from frustrated Kitaev model~\cite{Kitaev}.
This research has developed rapidly after a seminal paper of Jackeli and Khaliullin~\cite{Jackeli09} was published. They proposed that the QSL
might be realized in Mott insulators with strong spin-orbit coupling,
opening a new route to seek for the Kitaev QSLs.

In recent years it was recognized that $S=1$ variant could be designed
by considering strong Hund's coupling among two electrons in $e_g$
orbitals and strong spin-orbit coupling at anion sites~\cite{Sta19},
and this innovative concept sparks an intensive theoretical studies.
The 1D version of the Kitaev model can be considered as a limiting case
with vanishing interactions along $z$ bonds of hexagonal lattice. The
XX- and YY-type nearest-neighbor Ising interactions toggle sequentially
between odd and even bonds in the 1D Kitaev chain. Despite severe
simplification, the Kitaev chain shares some features of the honeycomb
model.

In this paper, we study the spin-1 Kitaev chain, which is nonintegrable.
We show that this model after increasing spin to $S=1$ surprisingly
harbors an extensive set of anomalous scarred eigenstates at finite
energy density that exhibit subextensive entanglement entropy. These
scarred states survive certain Heisenberg perturbation. Our results thus
firmly establish the existence of QMBSs in the 1D spin-1 Kitaev model.
The entanglement entropy of a subregion $A$ in an eigenstate $\alpha$ is,
\mbox{${\cal S}^\alpha_A=-{\rm Tr}\left\{\rho_A^\alpha\ln\rho_A^\alpha\right\}$,}
and ETH-obeying states have extensive volume-law entanglement entropy,
${\cal S}_A^\alpha\propto V$. Thus, to show that the quantum states
violate the ETH, we need only to show that their entanglement entropy
is subextensive.

The purpose of this paper is twofold: First, we would like to provide
evidence that scarred eigenstates indeed appear naturally in the spin-1
Kitaev model as the PXP model is embedded in one of subspaces. Second,
quantum scars in many-body Hamiltonian relevant for electronic systems
are also of interest. Thereby we are guided by the idea that studying a
wide variety of interacting systems with exact scar states and their
stability to perturbations would be beneficial for the general
understanding.

\section{Spin-1 Kitaev Chain}
\label{sec:chain}

In the present paper, we focus on a spin-1 Kitaev chain,
\begin{eqnarray}
\hat{H}_{K}&=&K\sum^{N/2}_{j=1}\left(
S^x_{2j-1}S^x_{2j}+S^y_{2j}S^y_{2j+1}\right)
\label{Ham1},
\end{eqnarray}
where spin operators $\{S_j^a\}$ (with $a$=$\{x,y,z\}$) are the spin-1
operators at site $j$, for a chain of $N$ sites. The spin operators obey
the SU(2) algebra, $[S_i^a,S_j^b]=i\delta_{ij}\epsilon_{abc}S_j^c$, with
the totally antisymmetric tensor $\epsilon_{abc}$ and
$({\bf S}_j)^2\!=S(S+1)\!=2$.
The spin-1 Kitaev model in Eq. (\ref{Ham1}) breaks the global spin
rotation SU(2) symmetry, while still obeys the time reversal symmetry
${\cal T}$, ($S_j^{x,y,z}\to -S_j^{x,y,z}$) and the spatial inversion
symmetry ${\cal I}$, ($j\to N-j+1$, {$S_j^{x,y,z}\to S_{N+1-j}^{x,y,z}$).
The sign of $K$ in Eq. (\ref{Ham1}) is irrelevant as far as a rotation
$U_z=\prod_j\Si_{2j}^z$ at each even lattice site by an angle $\pi$
about the $z$-axis will reverse the sign.

While the sign of the Kitaev
interactions is still under debate, with conflicting results from
theoretical and experimental studies, RuCl$_3$ was deemed as a promising
candidate for the Kitaev QSL with spin $S=\frac12$ \cite{Win17}. For the
present model Eq. \eqref{Ham1} we shall investigate whether QSL is also
stable in certain regime of parameters. Previous neutron scattering
studies performed to complement density functional theory (DFT) suggest
that the Kitaev interaction may be ferromagnetic (FM) or AFM. A recent
experiment on $\alpha$-RuCl$_3$ by measuring azimuthal dependence
\cite{Sears20} suggests that the Kitaev interaction is here~FM.

It is convenient to use a representation for $S=1$ given by the set of
orthonormal states $\{|-\rangle,|0\rangle,|+\rangle\}$, where
\begin{eqnarray}
\vert -\rangle\equiv&\frac{1}{\sqrt{2}}(\vert-1\rangle-\vert1\rangle),\nonumber \\
\vert +\rangle\equiv&\frac{i}{\sqrt{2}}(\vert-1\rangle+\vert1\rangle),
\end{eqnarray}
and $\vert m\rangle$ is an eigenstate of the spin operator $S^z_i$ to
an eigenvalue $m=-1, 0, 1$, i.e., it stands for the state
$\vert1,m\rangle$. \mbox{Furthermore,} spin-1 operators can be written
as $S^a_{bc}=i\ep_{abc}$. Explicitly, spin operators $\{S^x,S^y,S^z\}$
for spin $S=1$ may be represented by matrices
\begin{eqnarray}
%S^x&= &
\left(
\begin{array}{ccc}
0 & 0 & 0 \\
0 & 0 & -i \\
0 & i & 0 \\
\end{array}
\right),\quad
%S^y=
\left(
\begin{array}{ccc}
0 & 0 & i \\
0 & 0 & 0 \\
-i & 0 & 0 \\
\end{array}
\right),\quad
%S^z=
\left(
\begin{array}{ccc}
0 & -i & 0 \\
i & 0 & 0 \\
0 & 0 & 0 \\
\end{array}
\right).
\label{s1adef}
\end{eqnarray}
The spin components define site parity operators $\Si^a_j=e^{i\pi S_j^a}$,
i.e., $\{\Si^x,\Si^y,\Si^z\}$ are given by the matrices:
\begin{eqnarray}\label{sigx1}
\left( \begin{array}{ccc}
1 & 0& 0 \\
0 &-1& 0 \\
0 & 0&-1 \end{array} \right),\;
%\Si^y =
\left( \begin{array}{ccc}
-1& 0& 0 \\
 0& 1& 0 \\
 0& 0&-1 \end{array} \right),\; %\non \\
%\Si^z &=&
\left( \begin{array}{ccc}
-1& 0& 0 \\
 0&-1& 0 \\
 0& 0& 1 \end{array} \right).
\label{Sixyz} \end{eqnarray}
From the form of these operators we see that the eigenvalues of
operators $\Si^a_j$ are $\pm 1$, but $-1$ is doubly degenerate. The
Hamiltonian in Eq. (\ref{Ham1}) respects the dihedral group $D_2$ for
a global discrete symmetry with a rotation by an angle $\pi$ about the
$\{x,y,z\}$ axes, i.e., $\prod_j \Si_j^a$. One finds that all $\Si^a_j$
matrices commute with each other. In addition, $\Si^a_j$ commutes with
$S_j^a$ but anticommutes with $S_j^b$ ($a\!\neq\!b$), i.e.,
\mbox{$\{\Si^a_j,S_j^b\}=\{\exp(i\pi S_j^a),S_j^b\}=0$.}

In terms of the ladder operators $S_j^\pm\equiv S_j^x\pm i S_j^y$,
one finds that
\begin{eqnarray}
[S_j^+, S_j^-]&=&2S_j^z, \quad [S_j^z,S_j^\pm]=\pm S_j^\pm, \nonumber \\
\{S_j^+, S_j^-\}&=&2\left(2-(S_j^z)^2\right).
\end{eqnarray}
The Ising terms in Eq. (\ref{Ham1}) change the total pseudospin-$z$ at
both $x$-link $(2j$-$1,2j)$  and  $y$-link $(2j,2j$+$1)$ by either 0 or
$\pm 2$. Thus the bond parity operators on odd/even bonds,
\begin{eqnarray}
\hat{W}_{2j-1}=\Si^y_{2j-1} \Si^y_{2j},  ~~{\rm and}~~
\hat{W}_{2j}=\Si^x_{2j} \Si^x_{2j+1},
\label{originalW}
\end{eqnarray}
define the invariants of the Hamiltonian in Eq. (\ref{Ham1}).
The different forms on odd and even bonds in Eq. (\ref{originalW})
can be cured by a unitary transformation on the even sites,
\begin{equation}
\label{e_rot}
S^x_{2j} \to S^y_{2j}, ~~S^y_{2j} \to S^x_{2j}, ~~{\rm and} ~~S^z_{2j}
\to - S^z_{2j}. \end{equation}
In the transformed frame, the Hamiltonian in Eq. (\ref{Ham1}) then
takes a translationally invariant form,
\begin{eqnarray}
\tilde{H}_{\rm K}=K \sum_{j=1}^{N}~S^x_jS^y_{j+1}, \label{ham2}
\end{eqnarray}
and the bond parity operators on odd/even bonds then simultaneously take
a universal convenient form,
\begin{eqnarray}
\label{convenientWj}
\tilde{W}_{j}=\Si^y_{j} \Si^x_{j+1}.
\end{eqnarray}
The eigenvalues of $\tilde{W}_j$ are $w_j=\pm1$. One observes from Eq.
(\ref{sigx1}) that $\Si_j^z$= $\Si_j^x\Si_j^y$, implying two of them
are independent, e.g. $\Si_j^x$ and $\Si_j^y$. The $\mathbb{Z}_2$-valued
conserved quantity $\tilde{W}_{j}$ implies that the Hilbert space can be
decomposed into $2^N$ sectors of unequal sizes for spin-1 chain.

The projector at a subspace with a given set of invariants,
$\{w_1,w_2,\cdots, w_N\}$, can be constructed as
$\hat{Q}({w_1,w_2,\cdots, w_N})=\prod_j(1+ w_{j}\tilde{W}_{j})/2$. It
acts on a random state $\vert\psi_{\rm rand}\rangle$ and the dimension
$\cal{D}$ can be calculated by using the technique of transfer matrix.
The ground state of $\tilde{H}_{\rm K}$ with periodic boundary
conditions (PBCs) lies in the sector with defect-free subspace
\cite{You08}. Different from the spin-1/2 counterpart, where the uniform
subspace, with parity being either $w_j=+1$ or $w_j=-1$ for all $j$, are
all equivalent up to a unitary transformation, but the dimensions of
these sectors are nonidentical. The dimension of the sector with all
$w_j=-1$ is equal to ${\cal D}(\{-1,-1,\cdots,-1\})=2$, which appears to
be the smallest subspace, as is evidenced in Fig. \ref{SubspaceD_Kitaev}.

The two-dimensional subspace with all $w_j=-1$ can be rewritten as
$\vert ---\cdots --\rangle$ and $\vert +++\cdots ++\rangle$ defined in
the rotated bases in Eq. (\ref{e_rot}). On the contrary, the dimension
$\cal{D}$ for the subspace with parity $w_j=+1$ for all $j$ is proven
to be the Lucas number, i.e.,
${\cal D}(\{1,1,\cdots,1\})=F_{N-1}+F_{N+1}$, where $F_n$ is the
$n$-th Fibonacci number. More precisely,
${\cal D}(\{1,1,\cdots,1\})=g^N+g^{-N}$, where
\mbox{$g=(1+\sqrt{5})/2\approx 1.618$} is the golden ratio. It is
interesting to observe that such fractal dimension-based subspace is
the largest among all the sectors, see Fig. \ref{SubspaceD_Kitaev}.

\begin{figure}[t!]
\includegraphics[width=\columnwidth]{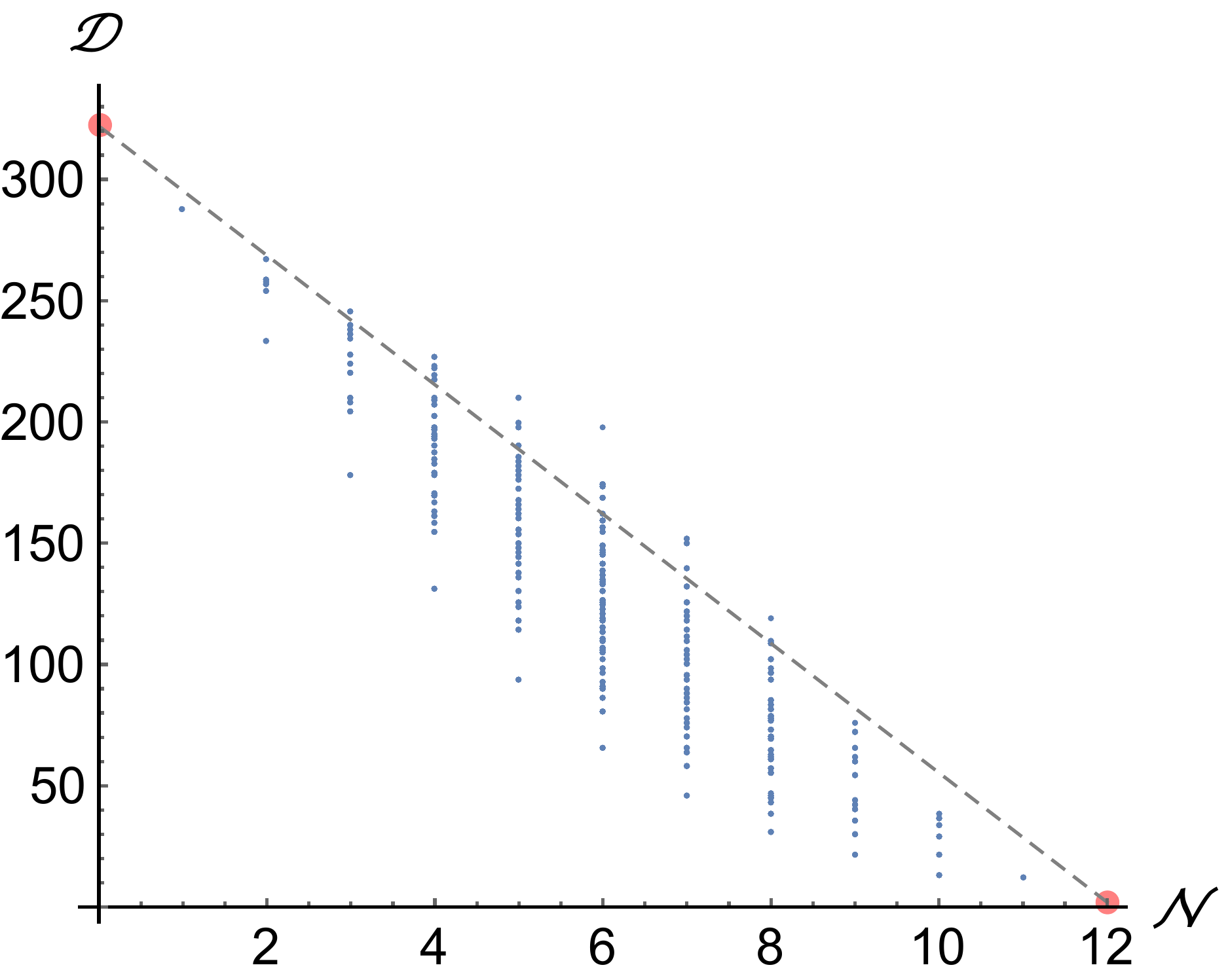}
\caption{The dimension of subspaces ${\cal D}$ for given parity
$(w_1,w_2,\cdots,w_N)$ as a function of the defect number
\mbox{${\cal N}=N-\sum_{j=1}^N w_j$,} for a chain with $N=12$ sites.
The red dot denotes the maximum (minimum) of $\cal{D}$ corresponding to
the uniform subspace with parity $w_j=+1$ ($w_j=-1$) for all $j$.
The dashed line is a guide for the eye. }
\label{SubspaceD_Kitaev}
\end{figure}

The constrained Hilbert space without any adjacent spin-up atoms makes
the system identical to that of chains of \mbox{spin-1/2} atoms. Hence,
the Hilbert space of a given sector can be mapped into the Hilbert
space of a spin-1/2 chain with some states excluded. Accepting the
notation that the $\vert{\uparrow}\rangle$ state represents the state
$\vert -\rangle$ (head), and the $\vert{\downarrow}\rangle$ state
stands for either the state $\vert 0\rangle$ (empty) or the state
$\vert +\rangle$ (tail), the system can be mapped to a single qubit-flip
model with nearest-neighbor exclusion represented by the effective
Hamiltonian in the sector with $w_j=+1$ for all $j$~\cite{Sen10}:
\begin{equation}
  \label{ham3}
{\cal H}_{\{1, 1,\cdots, 1\}}~=~ \sum_{j=1}^N~ \tilde{X}_j.
  \end{equation}
Here $\tilde{X}_j\equiv P_{j-1}X_jP_{j+1}$, where Pauli operators
\mbox{$X_j=\left(\,\vert{\uparrow}\rangle\langle{\downarrow}\vert
+\vert{\downarrow}\rangle\langle{\uparrow}\vert\,\right)$,}
$Z_j=\left(\,\vert({\uparrow}\rangle\langle{\uparrow}\vert
-\vert{\downarrow}\rangle\langle{\downarrow}\vert\,\right)$,
\mbox{and the} projectors
$P_j=\vert{\downarrow}\rangle\langle{\downarrow}\vert=(1-Z_j)/{2}$,
ensure that the nearby atoms are not \textit{simultaneously} in the
excited state. The corresponding dimension of the constrained Hilbert
space is exactly equal to ${\cal D}(\{1,1,\cdots,1\})$ through an
isomorphic mapping. Note that the PXP model is an effective model of
biaxial Ising model with both transverse and longitudinal fields, where
the low field $h_x\ll J_z$, and the saturation field $h_z=2J_z$ are
present. In spite of its rather friendly form, the Hamiltonian Eq.
(\ref{ham3}) is nonintegrable due to the low-energy constraint imposed
on the Hilbert space.

\begin{figure}[t!]
\includegraphics[width=\columnwidth]{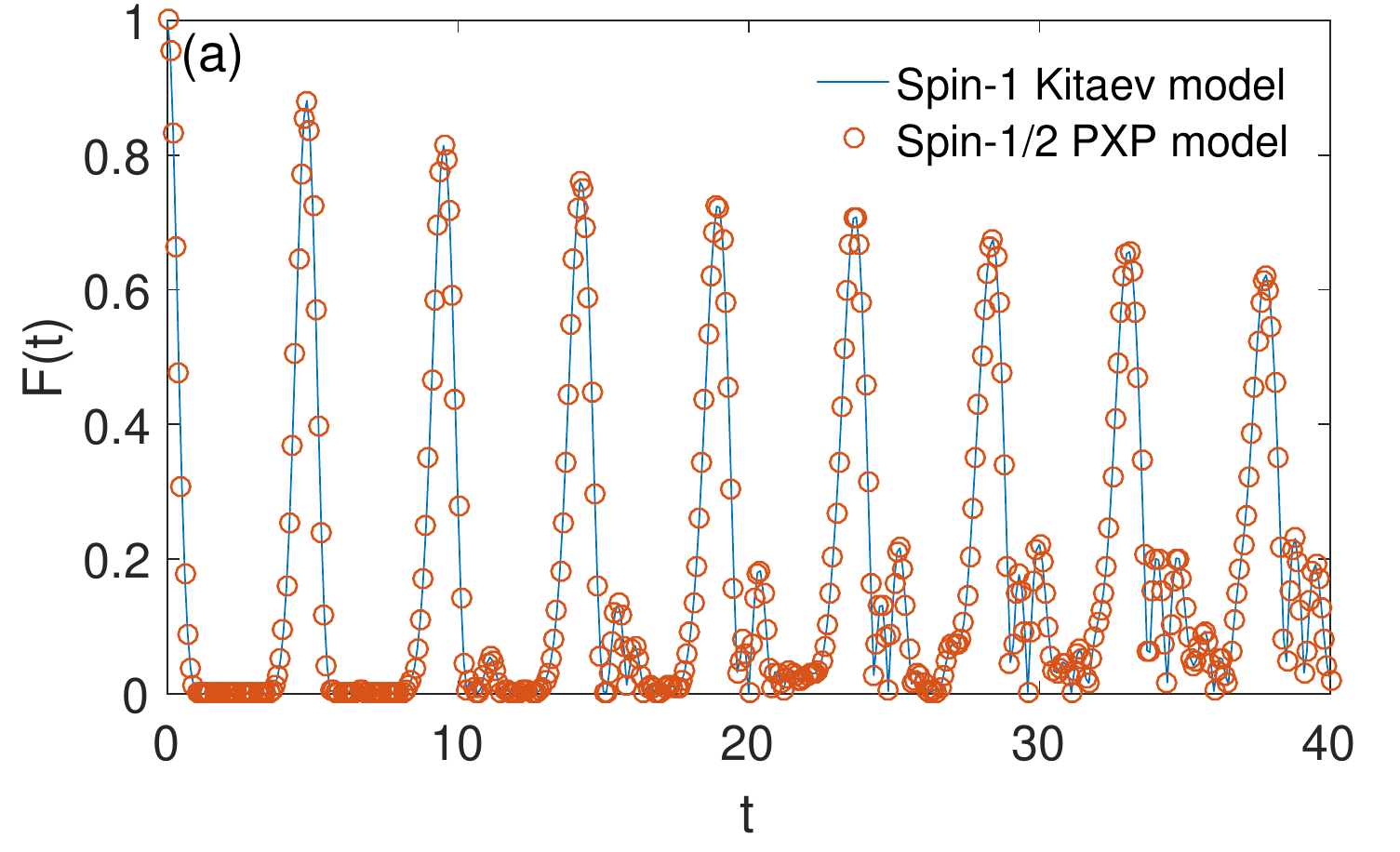}
\includegraphics[width=\columnwidth]{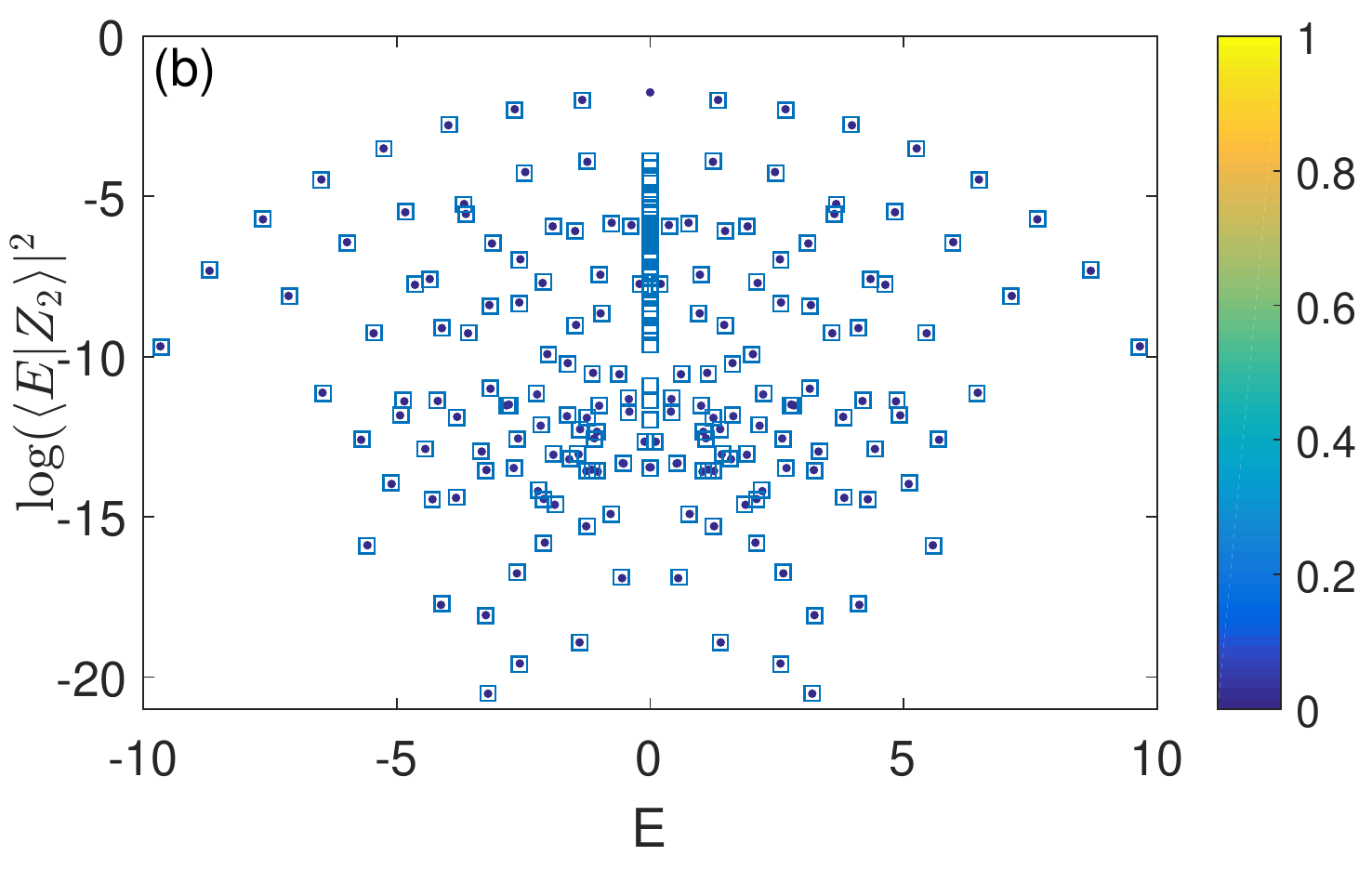}
\caption{Characteristic quantum features of the spin-1/2 PXP model:
(a)~The fidelity $F(t)$ for spin-1 Kitaev model starting from the
initial $|\tilde{\mathbb{Z}}_2\rangle$ ($|\mathbb{Z}_2\rangle$)
states with $N=18$;
(b) Eigenstate overlap with the N\'{e}el $\mathbb{Z}_2$ state for
spin-1/2 PXP model (dots) and the $w_j=1$ for $(j=1,2,\cdots,N)$
subspace of spin-1 Kitaev model (squares) with $N=16$. Color scale
on the right to panel (b) indicates the density of data points,
with lighter regions being more dense.
}
\label{F_Z2_both_N16}
\end{figure}

It has been noted that PXP model specified by the Hamiltonian
(\ref{ham3}) will exhibit QMBSs, which can be experimentally prepared
and probed using a global quench. Experiment~\cite{Bernien17} and
numerical simulations on small systems~\cite{Sun2008} revealed that the
relaxation under unitary dynamics strongly depends on the initial state
of the system. It is found that special eigenstates have anomalously
high overlaps with certain product states.

The polarized state with all spins in $|{\downarrow}\rangle$ state,
\mbox{$\vert\varnothing\rangle=\prod_jP_j\vert\psi_{\rm rand}\rangle$,}
satisfies ${\cal H}_{\{1,1,\cdots,1\}}\vert\varnothing\rangle=0$, which
allows one to identify a dominant subset of special states in the PXP
model. We initialize the system at time $t=0$ in the state
$\vert\psi(0)\rangle\equiv\vert\mathbb{Z}_k\rangle$, namely,
\begin{equation}\label{eq:cdw}
|\mathbb{Z}_k\rangle=
\ldots\tilde{X}_{k}\ldots\tilde{X}_{2 k}\ldots|\varnothing\rangle,
\end{equation}
and then follow the evolution of the initial state with the PXP
Hamiltonian,
$\vert\psi(t)\rangle=
\exp\left(-i{\cal H}_{\{1,1,\cdots,1\}}t\right)\vert\psi(0)\rangle$.
The evolution is determined by the decomposition of
$\vert\psi(0)\rangle$ in terms of the eigenstates of
${\cal H}_{\{1,1,\cdots,1\}}$. The initial states with atoms in all
spin-down state $\vert\mathbb{Z}_1\rangle\equiv\vert\varnothing\rangle$,
or $|\mathbb{Z}_4\rangle$ show fast relaxation and no revivals,
characteristic of thermalizing systems. Remarkably, the quantum quench
from either period-2 state ($|\mathbb{Z}_2\rangle$) or the period-3
state ($|\mathbb{Z}_3\rangle$) will surprisingly give rise to coherent
oscillations, which can be observed in the dynamics by measuring the
expectation values of certain local observables and the quantum fidelity,
\begin{eqnarray}
\label{Ft}
F(t)=|\langle\mathbb{Z}_k|\exp(-iHt)|\mathbb{Z}_k\rangle|.
\end{eqnarray}
The observed oscillations and the apparent non-ergodic dynamics are
attributed to the existence of QMBSs \cite{Ser21}.

Intuitively, the empty state $\vert\varnothing\rangle$ corresponds to
the state $\vert 0000\ldots 00\rangle$, which can be inferred from Eq.
(\ref{sigx1}) that the configuration $\vert 0000\ldots 00\rangle$
resides in this subspace with uniform $\mathbb{Z}_2$ invariants, i.e.,
$w_j=1$ for all $j$. In this context, the akin states
$|\tilde{\mathbb{Z}}_k\rangle$ of spin-1 Kitaev model Eq. (\ref{ham2})
as product states in Eq. (\ref{eq:cdw}), can be reconciled by acting
\begin{equation}
\label{eq:cdw2}
|\tilde{\mathbb{Z}}_k\rangle=\ldots S^x_{k}S^y_{k+1}\ldots
S^x_{2 k}S^y_{2k+1}\ldots\vert 0000 \cdots 00\rangle.
\end{equation}
Then the corresponding spin-1 states are given by:
\begin{eqnarray}
\label{Eq:Z2generate2}
|\tilde{\mathbb{Z}}_2\rangle&=& \vert -+-+ \cdots -+  \rangle, \\
|\tilde{\mathbb{Z}}_3\rangle&=& \vert +-0+-0\cdots +-0\rangle, \\
|\tilde{\mathbb{Z}}_4\rangle&=& \vert +-00+-00 \cdots +-00 \rangle.
\end{eqnarray}
It is clear that the translated states,
$|\tilde{\mathbb{Z}}'_k\rangle=(T_{i\to i+1})^j\,|\mathbb{Z}_k\rangle$
\mbox{($j=1,2\cdots,k-1)$}, also belong to the QMBSs, where
$T_{i\to i+1}$ denotes the translation by one lattice site, i.e.,
\mbox{$i\to i+1$}. As is shown in Fig. \ref{F_Z2_both_N16}(a), the
revivals of the quantum fidelity for spin-1 Kitaev model (\ref{ham2})
[spin-1/2 PXP model (\ref{ham3})] starting from the initial N\'{e}el
state $|\tilde{\mathbb{Z}}_2\rangle$ ($|\mathbb{Z}_2\rangle$) are
indistinguishable, suggesting the existence of ETH-violating QMBSs.

\begin{figure}[t!]
\includegraphics[width=\columnwidth]{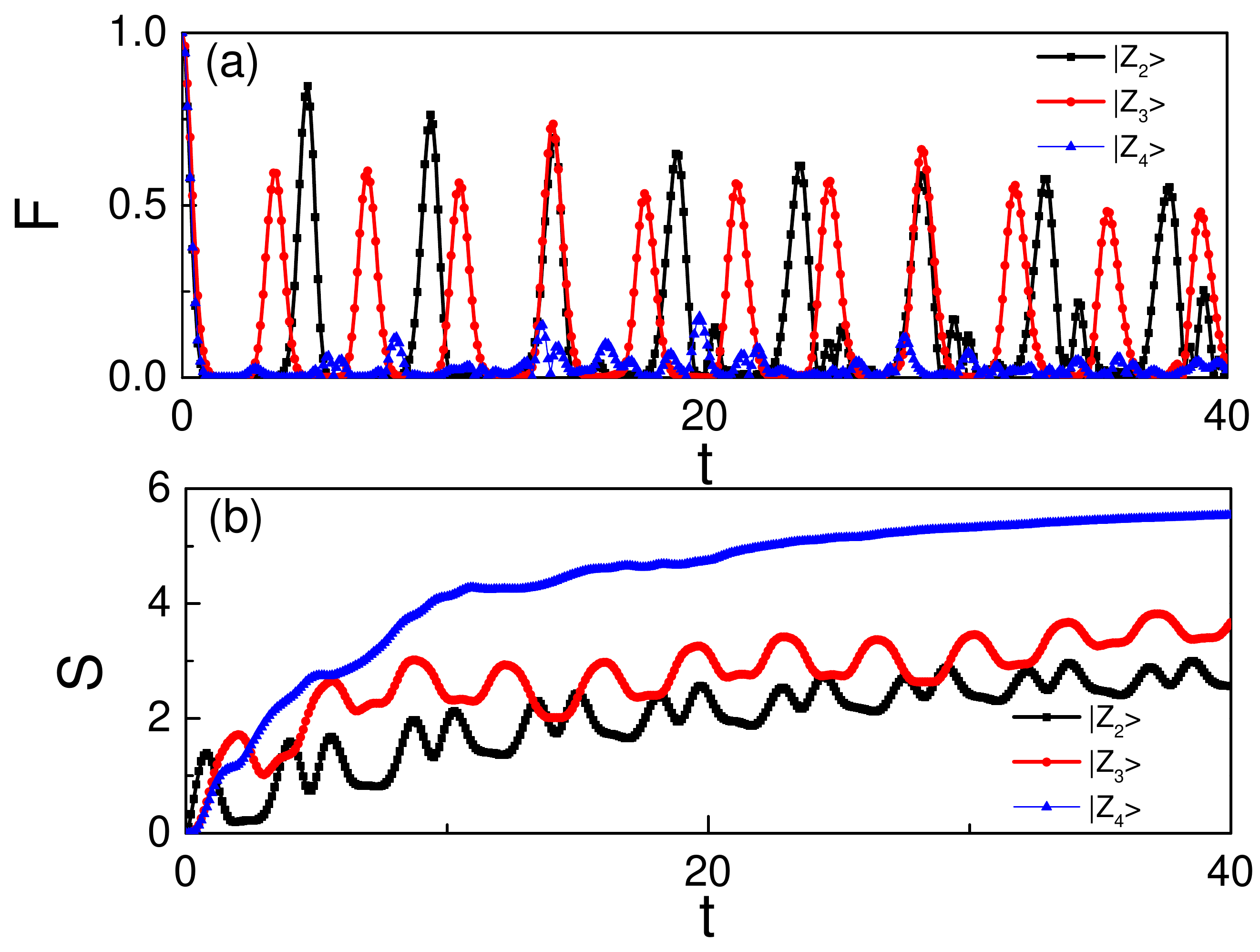}
\caption{Quantum features of the 24-site $S=1$ Kitaev chain:
\mbox{(a) the fidelity $F$ and (b) the entanglement ${\cal S}$.}
The results correspond to starting from the initial
$|\tilde{\mathbb{Z}}_k\rangle$ ($k$=2,3,4) states.
The bond dimension is set as $\chi=300$.}
\label{F_S_spin1_Zn_N24}
\end{figure}

Figure \ref{F_S_spin1_Zn_N24} shows that the fidelity $F$ and
entanglement ${\cal S}$ with its dynamics for different initial
states are interrelated in a subtle way.
One can observe the coherent oscillations of fidelity that persist
for long times for quenches from initial product states
$|\tilde{\mathbb{Z}}_2\rangle$ and $|\tilde{\mathbb{Z}}_3\rangle$
\cite{Tur182}, with different revival periods, and the entanglement
entropy ${\cal S}$, see Fig. \ref{F_S_spin1_Zn_N24}(b),
is gradually growing with time. The weak oscillations of the
entanglement are clearly visible by eliminating the linear growth,
featuring the many-body revivals~\cite{Turner18}.

By contrast, the fast damping amplitude of fidelity for quench from
$|\tilde{\mathbb{Z}}_4\rangle$ state indicates the negligible overlaps
between scarred eigenstates and the $|\tilde{\mathbb{Z}}_4\rangle$
product state. Meanwhile, the entanglement undergoes an extremely fast
growth. Note that the bipartite entanglement approaching the maximum
entanglement entropy $\log_2(\chi d)$ implies a large accumulated error
and thus an inevitable breakdown of the density-matrix renormalization
group (DMRG) method~\cite{Gobert05}.

Furthermore, it is striking to find that the energy spectra and
eigenstates overlap with the N\'{e}el state of Eq. (\ref{ham3}) also
coincide with those of Eq. (\ref{ham2}), except the highly degenerate
states at $E=0$, implying that a precise relation exists between the
eigenstate of spin-1 Kitaev model (\ref{ham2}) and the celebrated
spin-1/2 PXP model (\ref{ham3}). The zero modes of the PXP model at
the center of the spectrum are pinned by the particle-hole symmetry,
i.e., $\{{\cal C},H_{\rm PXP}\}=0$, with
${\cal C}=\prod_{j=1}^N Z_j$~\cite{Lin19}. The states with $E=0$ can
be classified as eigenstates of ${\cal C}$. An exponential growth of
protected many-body zero-energy modes with system size arise as a
consequence of intertwining of spectral-reflection ${\cal C}$
symmetries and relevant point-group symmetries~\cite{Schecter18}.

Among the above macroscopic number of states, two exact scar states for
PBC have been identified in terms of matrix product states with
a finite bond dimension~\cite{Lin19}, which shows that these exact scar
states have constant entanglement. However, the bipartite
entropy of the eigenstates of spin-1/2 PXP model in Fig.
\ref{S_both_N16}(a) and the $w_j=1$ \mbox{$(j=1,2,\cdots,N)$} subspace
of spin-1 Kitaev model in Fig. \ref{S_both_N16}(b) quantitatively
exhibit similar features but are not exactly equivalent.
Notably, the different choice of reference bases will modify the values
of bipartite entanglement entropy.

\begin{figure}[t!]
\includegraphics[width=\columnwidth]{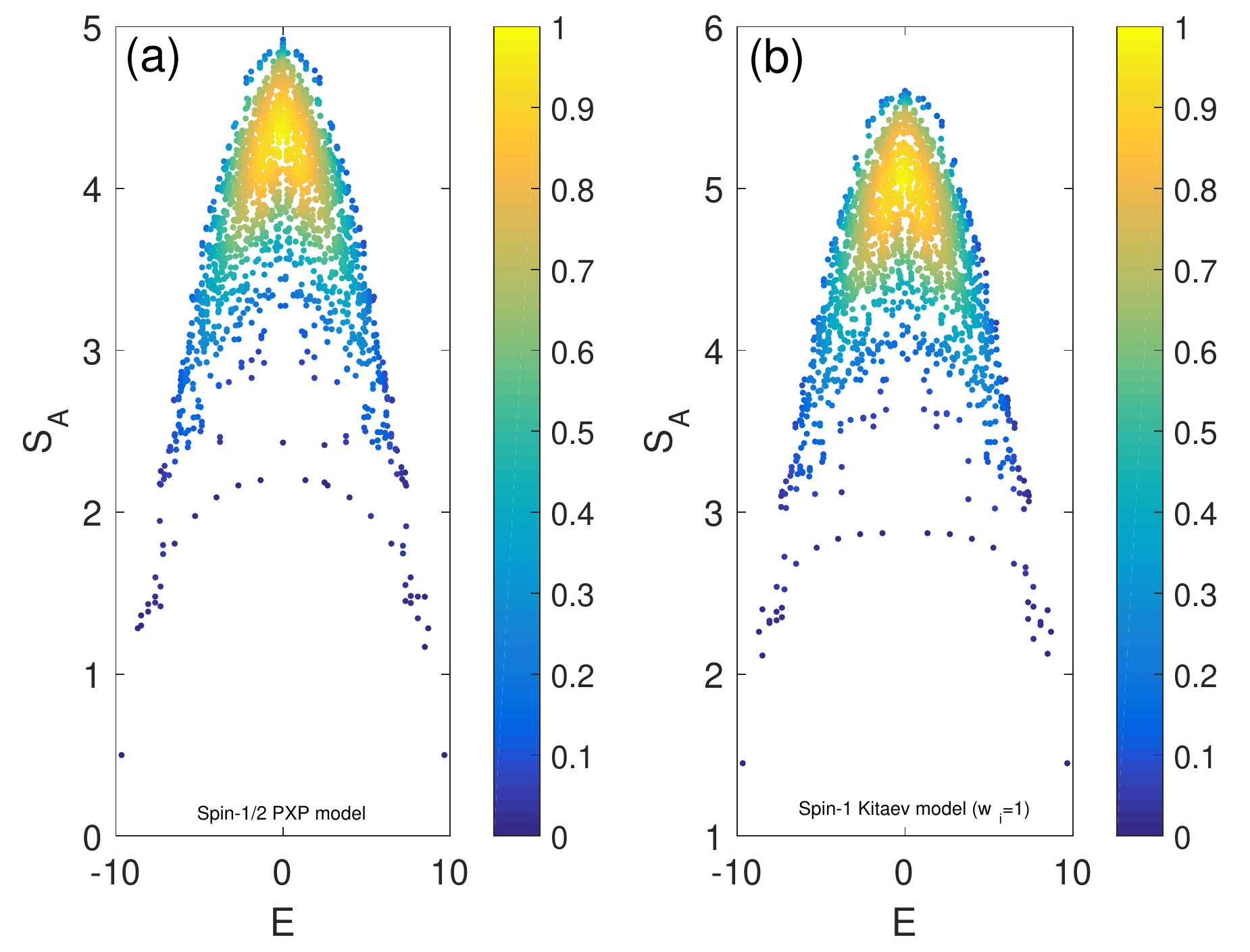}
\caption{Bipartite entropy of the eigenstates for:
(a) spin-1/2 PXP model and
(b) spin-1 Kitaev model, within $w_i=1$ \mbox{$(i=1,2,\cdots,N)$}
subspace. Color scale indicates the density of data points,
with lighter regions being more dense. Here the chain has $N=16$ sites.
}
\label{S_both_N16}
\end{figure}

The explicit technical insights can
be accessible by an exact solution of the two-site model, which are
presented in the Appendix \ref{bipartiteentropy_N2}. The scarred
eigenstates are characterized by the low bipartite entanglement entropy
${\cal S}_A$, and they become increasingly decoupled from the thermal
bulk states as the system size $N$ increases. The primary
$\mathbb{Z}_2$ scar states can be constructed from the excitations on
top of the exact scar states at $E=0$. The equally-spaced towers of
QMBSs in the spectrum are closely related to perfect revivals, but
these characteristics of particular interest are currently limited by
small system sizes. Yet, they are experimentally relevant.

\section{Stability of QMBS against Heisenberg interactions}
\label{sec:sta}

It is appealing to consider the stability of the QMBS to a nonzero
perturbation, which could possibly preserve or undermine the
conservation of local quantities characteristic of the Kitaev model.
Considering atomic states in the $t_{2g}$ manifold are immersed under
the crystalline electric field with the strong spin-orbit coupling,
other interactions, such as the Heisenberg exchange,
and Dzyaloshinskii-Moriya interaction, coexist with the Kitaev
interactions~\cite{Sano18}. In addition to the quadratic exchange
interactions, the uniaxial single-ion anisotropy (SIA) due to
crystal-field effects~\cite{Xu18} and further-neighbor interactions
are also ubiquitous~\cite{Pesin10}. In reality, these interactions
play an important role in establishing non-trivial electronic
correlations for low-dimensional models with higher spin $S>\frac12$.

We are interested in effects induced by perturbations that leave
the intermediate symmetries intact. The SIA,
\begin{eqnarray}
\hat{H}_{\rm D}=D \sum_j \left(S_{j}^z\right)^2,
 \label{HamD}
\end{eqnarray}
is one of them and leads to a quantum phase transition between a
topologically trivial phase and a nontrivial phase in spin-1 AFM
Heisenberg chain as predicted by Haldane
\mbox{\cite{Haldane1,Haldane2}}. Simultaneously, the SIA leads to a
low-entropy spin Mott insulator in the FM analogy~\cite{Chung21,Altman03}.
It is easy to find that Eq. (\ref{HamD}) is invariant under the
rotation (\ref{e_rot}) and commutes with local parity operators defined
in Eq. (\ref{convenientWj}). We can observe in Fig. \ref{F_OP_t_D2}
that periodic oscillations in the order parameter \cite{Schecter19},
\begin{equation}
\label{O1}
 O_1=\left\langle\left(S_1^+\right)^2\right\rangle,
\end{equation}
and the quench fidelity persist for long times when starting from
$|\tilde{\mathbb{Z}}_2\rangle$. As $D$ increases, the oscillation
period gets shorter.

\begin{figure}[t!]
\includegraphics[width=\columnwidth]{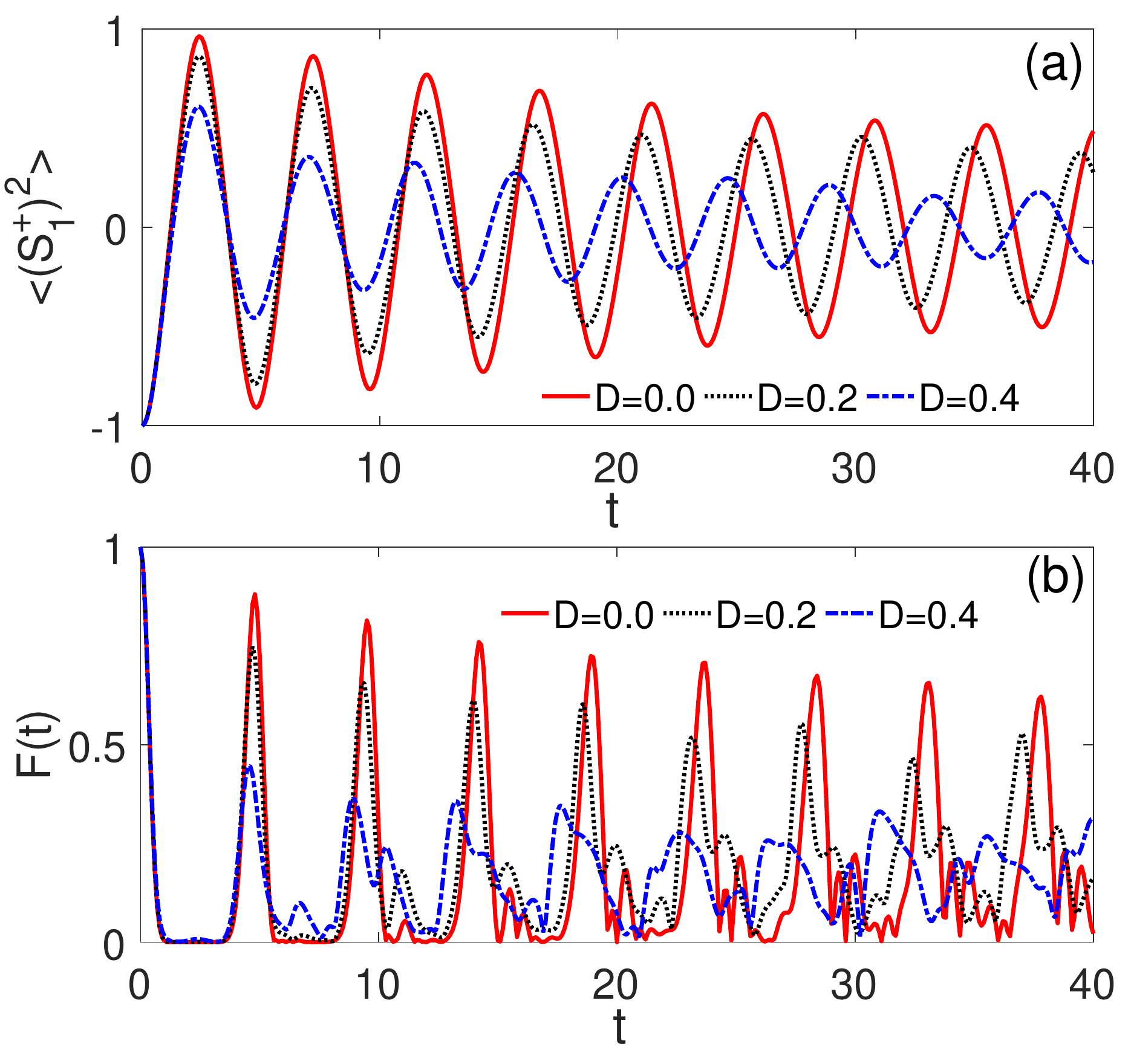}
\caption{Time evolution of an $18$-site Kitaev chain starting from the
initial $|\tilde{\mathbb{Z}}_2\rangle$ state:
(a) the order parameter \eqref{O1}, and
(b) the quench fidelity \eqref{Ft}.
The data are presented for increasing strength of single-ion anisotropy
\eqref{HamD}, $D=0.0$, 0.2, 0.4. see legend.}
\label{F_OP_t_D2}
\end{figure}

We are also interested in effects under a perturbation
\mbox{$V\propto (S_{j}^x S_{j+1}^z S_{j+2}^y$)} that obeys the same
intermediate $\mathbb{Z}_2$ symmetries of Eq. (\ref{ham2}). The
origin of such three-site XZY-type interactions is explained in Appendix
\ref{Kitaev_XZY}. One can observe that the order
parameter $O_1$ and the quench fidelity $F(t)$ show quite similar
oscillations as those of SIA. The state coherence sustains a long time
for a moderate perturbation, and the revival will degrade quantitatively
with increasing amplitude of multispin perturbations, while the
oscillation period is almost unchanged.

However, it is gradually recognized that isotropic Heisenberg
interactions commonly exist in solid state materials and perturbs the
systems with frustrated interactions. One frequently considered example
is the compass model where Heisenberg interactions are responsible for
the onset of possible long range 2D order in the ground state but
simultaneously excited states preserve their unique nature which may be
used for information storage. Here the nematic order which persists in
excited states may be used for correcting the faults along the
computations \cite{Tro12}.

Another example of robust structure of excited states in presence of
perturbation is encountered in the Kitaev-Heisenberg model for $S=1/2$
spin which has been intensely investigated and several phases with
broken symmetry have been found \cite{Win17,Chaloupka2010,Got17}.
A similar situation was reported as well for spin-1 Kitaev honeycomb
model in candidate materials, such as honeycomb Ni oxides with heavy
elements of Bi and Sb, where Kitaev interaction
is accompanied by a finite FM Heisenberg interaction. In the zero-field
limit, the Kitaev QSL is destabilized when $J/K>0.08$~\cite{Hickey20}.
We thus assume that such interactions are of Heisenberg type,
\begin{eqnarray}
 \hat{H}_{\rm J}=J \sum^{N}_{i=1} {\bf S}_i \cdot {\bf S}_{i+1},
 \label{Ham2}
\end{eqnarray}
where $J$ stands for the Heisenberg exchange coupling. After the
rotation (\ref{e_rot}), the perturbation by Heisenberg interaction can
be rewritten as
\begin{eqnarray}
\tilde{H}_{\rm J}=
J\sum^{N}_{j=1}\left(S_j^x S_{j+1}^y+S_j^yS_{j+1}^x-S_j^z S_{j+1}^z\right). \quad
\label{Ham6}
\end{eqnarray}
The full Hamiltonian takes an anisotropic translational form:
\begin{eqnarray}
\tilde{H}_{\rm KJ}\!=\sum^{N}_{j=1}\!
\left\{(K\!+\!J)S_j^xS_{j+1}^y\!+J S_j^yS_{j+1}^x\!-J S_j^zS_{j+1}^z\right\}\!. \quad
\label{Ham7}
\end{eqnarray}

A direct consequence is the additional Heisenberg interactions spoil the
$\mathbb{Z}_2$ symmetry of Eq. (\ref{Ham1}) associated with each bond.
It is worth noting that the energy levels of low-lying excited states
cross at $J=0$. The first excited state of $\tilde{H}_{\rm K}$ is
$N$-fold degenerate, corresponding to one $w_j=-1$ defect in the sector
with all other $w_j=1$, such that either
$\vert 00\!+\!0\ldots 00\rangle$ or $\vert 00\!-\!0\ldots00\rangle$
state occurs. A narrow gapped Kitaev phase is capable of sustaining the
perturbations of Heisenberg interactions.

The competition between spin-1 AFM Kitaev chain and isotropic
Heisenberg interactions has been analyzed for \mbox{$K=1$}
\cite{You20a}, while a complete phase diagram, to the best of our
knowledge, is elusive and deserves still a careful investigation.
To incorporate the phase diagram of spin-$1$ Kitaev-Heisenberg model
with $K=1$~\cite{You20a}, the ground state phase diagram with $K=-1$ is
depicted using the DMRG and exact diagonalization methods. In the DMRG
simulations, we keep up to $\chi=500$ bond states during the procedure
of basis truncation and the number of sweeps is $n=30$. These
conditions guarantee that the simulation is converged sufficiently fast
and the truncation error is smaller than $10^{-7}$.

\begin{figure}[t!]
\includegraphics[width=\columnwidth]{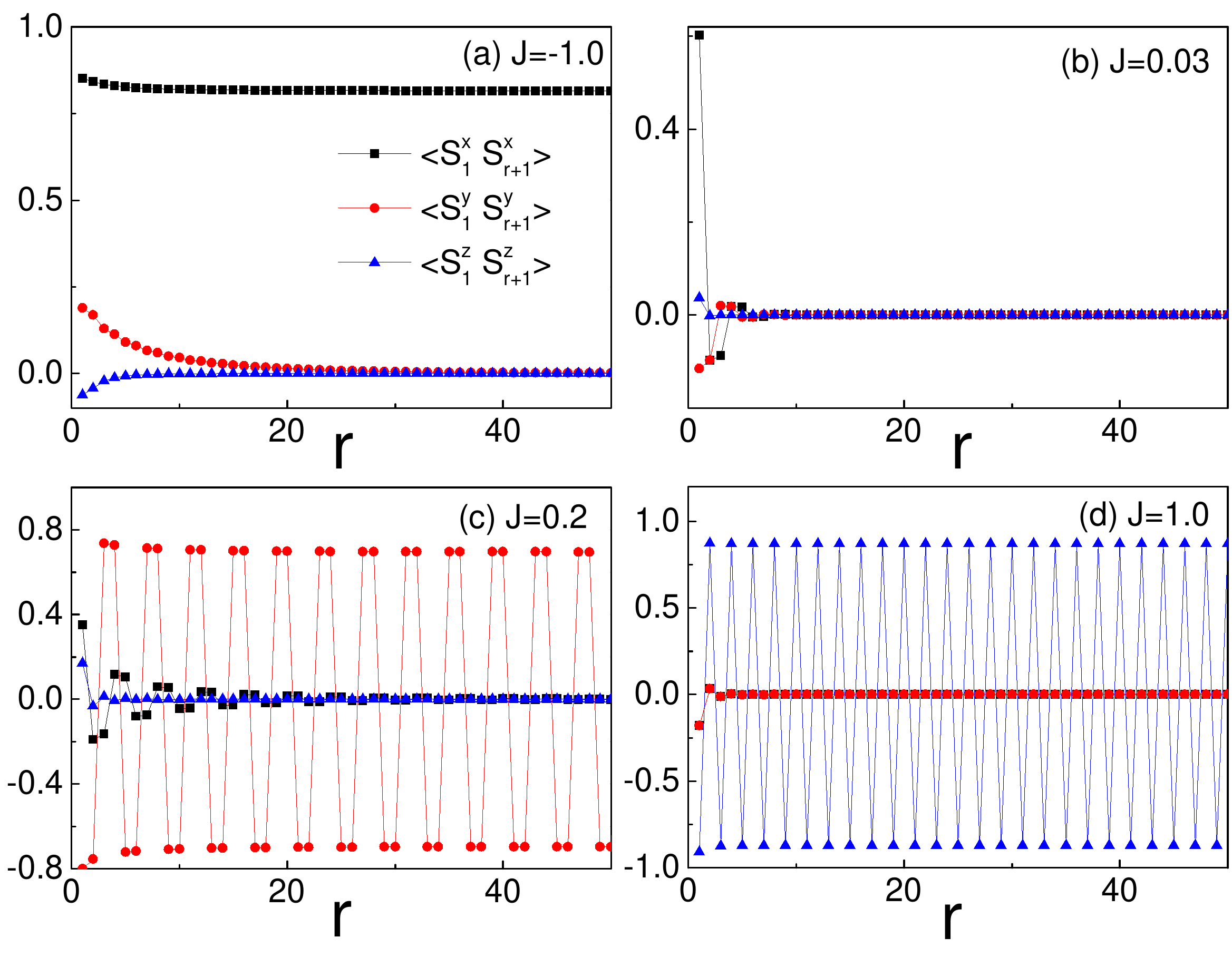}
\caption{The correlations between site 1 and $(r+1)$ for increasing
distance $r$ and for:
(a) $J=-1.0$, (b) $J=0.03$, (c) $J=0.2$, and (d)~$J=1.0$.
Here we use PBCs for the chain with $N=200$ sites.	
}
\label{CC_N200}
\end{figure}

One motivation for the study of AFM spin-1 Heisenberg chains
($K=0$, $J>0$) is that the model hosts a symmetry-protected topological
ground state with a finite Haldane gap \cite{White93}, a reminiscent of
AKLT state \cite{Affleck87}. It has been known for decades
that the indicator of the Haldane phase is beyond any local
symmetry-breaking order parameters. Instead, it can be characterized by
the exponential decay of two-spin correlation function and nonlocal
string order parameters~\cite{Nijs,Tasaki}.

As for the coexistence the Kitaev interaction and the Heisenberg
interactions, the spatial inversion symmetry, the time-reversal
symmetry, and the dihedral $D_2$ symmetry are preserved, which protects
the Haldane phase \cite{Pol10,Pol12}. Another specific case occurs with
$J\equiv -K/2$. Then Eq. (\ref{Ham7}) can be recast into an isotropic
form,
\begin{eqnarray}
\tilde{H}_{\rm KJ}&=&\frac{1}{2}K\sum^{N}_{j=1}
\left(\tilde{S}_j^x \tilde{S}_{j+1}^x+\tilde{S}_j^y\tilde{S}_{j+1}^y
+\tilde{S}_j^z \tilde{S}_{j+1}^z\right),
\label{Ham8}
\end{eqnarray}
through the following spin rotation
\begin{eqnarray}
\tilde{S}_j^x&=&\cos\left(\pi j/2\right)S_j^x -i\sin\left(\pi j/2\right)S_j^y, \nonumber \\
\tilde{S}_j^y&=&\cos\left(\pi j/2\right)S_j^y +i\sin\left(\pi j/2\right)S_j^x.
\end{eqnarray}
In this case, the ground state of Eq. (\ref{Ham8}) possesses one
Goldstone mode, and its character depends on the sign of $K$. The
Goldstone mode has a quadratic dispersion for FM Heisenberg model with
$K=-1$, while it is gapped for AFM Heisenberg model with $K=1$.

For a generic case, a well defined order parameter is a vital ingredient
for characterizing the nature of phases. In order to identify the region
of the Kitaev phase and the Haldane phase, we calculate the nonlocal
correlator
\begin{equation}
\label{Osigma}
C^a_{i,j}=\left\langle S^a_i\exp\left(i\theta\sum_{l=i+1}^{j-1}S_l^a\right)
S^a_j\right\rangle, ~~ a=x, y, z.
\end{equation}
For $\theta=\pi$, Eq. (\ref{Osigma}) becomes the den Nijs-Rommelse
string order parameter, whose limiting value
\mbox{$O^{a}_s={\rm lim}_{|i-j|\rightarrow\infty}\left(-C^{a}_{i,j}\right)$}
reveals the hidden symmetry breaking~\cite{Kennedy,Takada}.
For $\theta=0$, Eq. (\ref{Osigma}) reduces to two-point correlations.

\begin{figure}[t!]
 \includegraphics[width=\columnwidth]{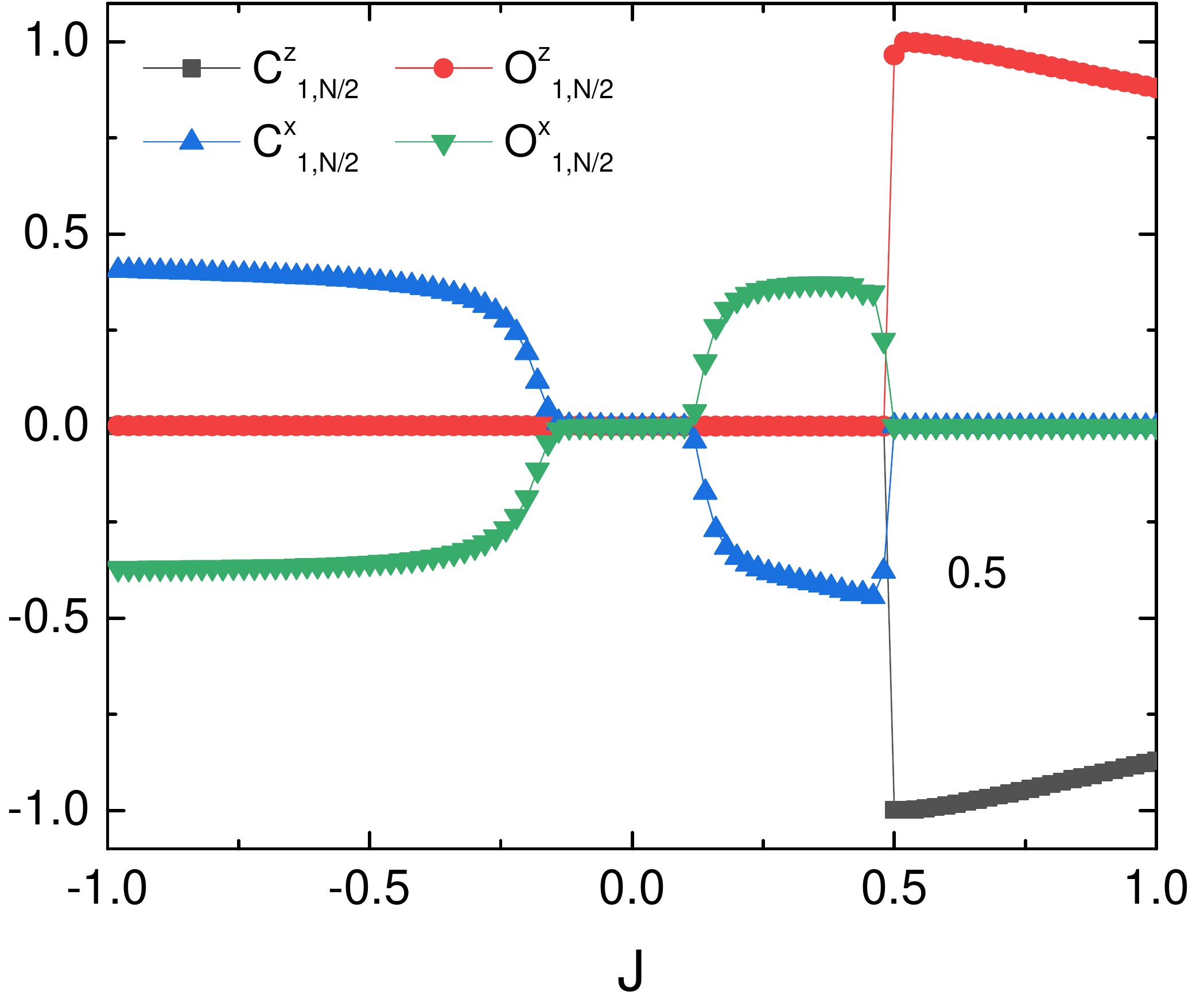}
\caption{Two-point correlation $C^a$ \eqref{Osigma} and string order
parameter $O^a$ ($a=x,z$) of 1D pristine spin-1 Kitaev-Heisenberg model
for $J\in (-1,1)$ between sites 1 and $N/2$ for $N=80$. With increasing
$J$, the order in the ground state changes from FM$_x$ through QSL and
LLRR phase to the Haldane phase.
}
\label{SOP_N100}
\end{figure}

When $J$ varies for $K=-1$, the competing correlations will trigger
miscellaneous phase transitions. For $J\simeq -1$, the $C^x_{i,j}$
correlations dominate, suggesting that the ground state is FM$_x$, see
Fig. \ref{CC_N200}(a). For $J\approx0$, the pure spin-1 Kitaev chain
hosts nearest neighbor AFM orders due to the $\mathbb{Z}_2$ symmetry.
The especially short-range correlations are demonstrated for $J=0.03$.
The nearest neighbor spins favor here the FM alignment as the next
nearest neighbor interactions are FM, this state is shown in Fig.
\ref{CC_N200}(b).
When $J$ increases further, the dominating $y$-component correlations
have a negative sign on odd bonds and a positive sign on even bonds.
These correlations indicate that spin order develops into the
left-left-right-right (LLRR) phase,
as is shown in Fig. \ref{CC_N200}(c).
Figure \ref{CC_N200}(d) shows that the $C^x_{i,j}$ correlations
dominate over $C^y_{i,j}$ and $C^z_{i,j}$ when $J\simeq 1$.

To clarify the
uniaxial order, the two-point correlation $C^a$ and the string order
parameter $O^a$ ($a=x$, $z$) between sites 1 and $N/2$ on a $N=80$ site
chain are shown for increasing $J$ in Fig. \ref{SOP_N100}. One can
recognize a finite $C^x_{1,N/2}$ that characterizes the FM$_x$ phase
for $J<-0.08$, while all correlations vanish in the interval
$-0.08<J<0.08$. Next $C^x_{1,N/2}<0$ for $0.08<J<0.50$,
while finally $O^z_{1,N/2}\simeq 0.9$ is finite for $J>0.5$
and defines the phase boundary of the Haldane phase.

\section{Phase diagram of the Kitaev-Heisenberg chain for $S=1$}

The parameters $\{K,J\}$ for the Kitaev and Heisenberg exchange
couplings can be set as
\begin{equation}
K \equiv\sin(\phi),  \quad\quad
J \equiv\cos(\phi),
\label{KJ}
\end{equation}
where the angle $\phi$ $\in[0,2\pi)$ parametrizes the Hamiltonian
\eqref{Ham7}. To this end, the complete phase diagram of the 1D spin-1
Kitaev-Heisenberg model as a function of $\phi$ is displayed
in Fig. \ref{PD_chart}, where the
vertical (horizontal) axis is the Kitaev $K$ (Heisenberg $J$) exchange
coupling. For both cases with either $K>0$ or $K<0$, the phase diagram
consists of four different phases, i.e., the FM phase, the Kitaev QSL
phase, the LLRR phase, and the Haldane phase. To some extent, the
symmetry between the regimes of $K>0$ and $K<0$ is restored.

\begin{figure}[t!]
\includegraphics[width=\columnwidth]{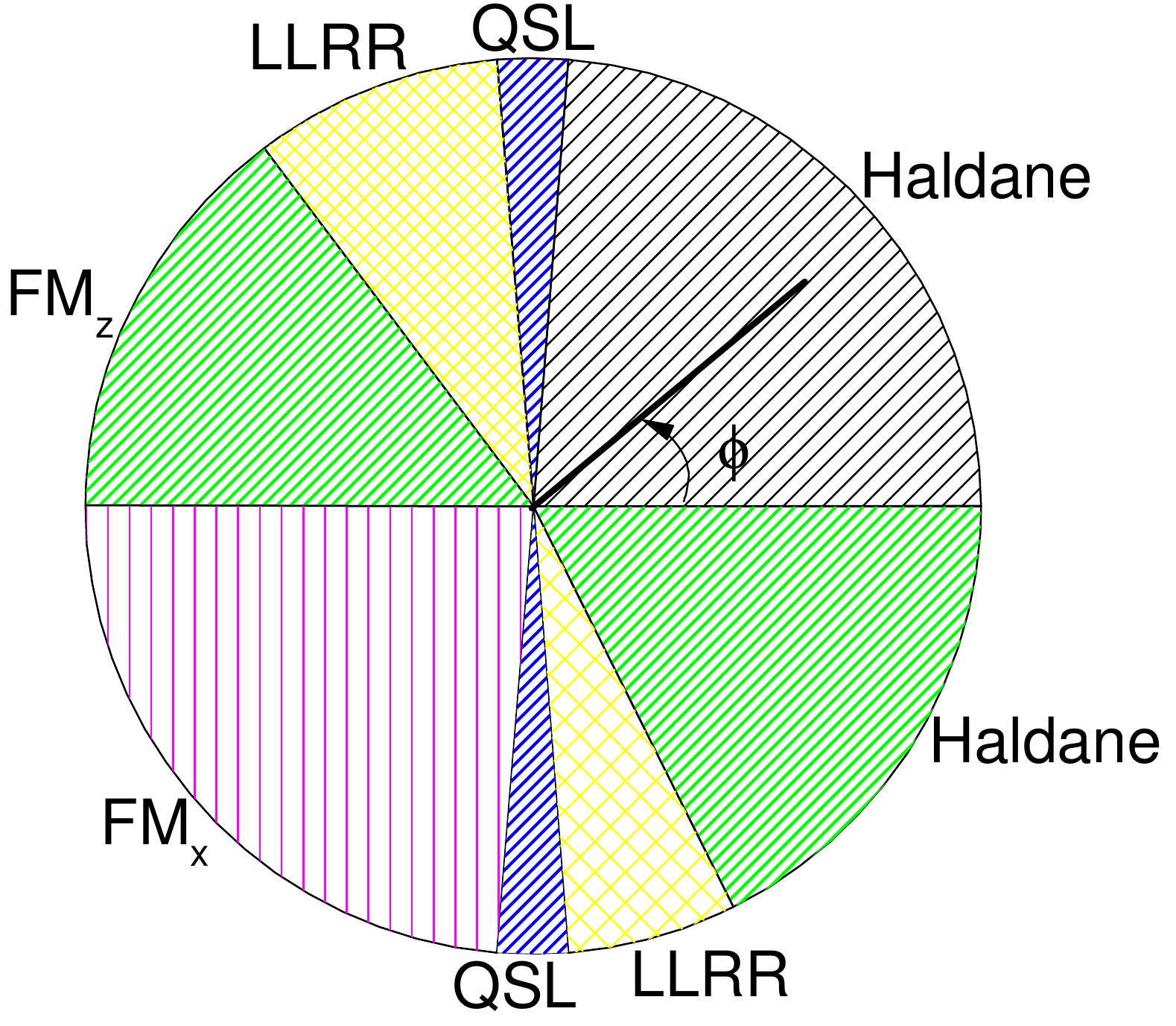}
\caption{Ground state phase diagram of the 1D Kitaev-Heisenberg model
for $S=1$. The vertical (horizontal) axis
%is
stands for
the Kitaev $K$  (Heisenberg $J$) exchange coupling, as given in Eq.
\eqref{KJ}. Here $\phi\in[0,2\pi)$ is the angle which determines the
exchange constants, see Eq. \eqref{KJ}.}
\label{PD_chart}
\end{figure}

It is interesting to note that a generalized Kennedy-Tasaki
transformation,
$U_{\mathrm{KT}}=\prod_{j<k}\exp{\left(i\pi S^y_jS^x_k\right)}$
\cite{Kennedy2,Oshikawa92}, can be realized as the reciprocal
transformation between the string order and the local FM order.
In this regard, Eqs. (\ref{Ham1}) and (\ref{Ham2}) are transformed into
equivalent ones with a minus sign standing with short-range
interactions, where the nonlocal string observable $O^a_{i,j}(\hat{H})$
is responsible for the two-point correlations $C^a_{i,j}(\tilde{H})$,
with $a=x,z$ of the transformed Hamiltonian,
\begin{eqnarray}\label{Ham10}
\tilde{H}^{\prime}_{\rm KJ}&=&
-K\sum^{N/2}_{j=1}\left(
S^x_{2j-1} S^x_{2j} + S^y_{2j} S^y_{2j+1}\right) \nonumber \\
&-&J\sum_{j=1}^N\left(
S_j^x S^x_{j+1}+  S_j^yS^y_{j+1} + \tilde{W}_{j} S_j^z S^z_{j+1}\right).
\end{eqnarray}
This suggests a likely relationship $\tilde{H}(-K,-J)\simeq\hat{H}(K,J)$.

Notably, the coupled $\mathbb{Z}_2$ gauge fields $\{\tilde{W}_{j}\}$ are
not conserved in the transformed pristine Hamiltonian. The duality
becomes exact when the strength of Heisenberg interaction $J$ is tiny
($\phi\approx\pm\pi/2$), which means that the parameters are within
the boundaries of the Kitaev QSL phases. However, Fig. \ref{PD_chart}
clearly shows that such a relationship does not hold for moderate values
of $J$ owing to quantum fluctuations of Ising gauge fields. The system
undergoes a second-order quantum phase transition from the FM phase to
the LLRR phase at \mbox{$\phi=0.8280\pi$} (i.e., for $K=1$, $J=-0.6$),
while another transition from the LLRR phase to the Haldane phase occurs
at $\phi=1.8524\pi$ \mbox{(i.e., for $K=-1$, $J=0.5$).} In fact, there
is no phase transition at $\phi=0$ ($K=0$, $J=1$), where the string
observables $O^a_{i,j}$ are balanced in both $x$- and $z$-directions,
corresponding to the restoration of
hidden $\mathbb{Z}_2\times\mathbb{Z}_2$ symmetry breaking.

\begin{figure}[t!]
\includegraphics[width=\columnwidth]{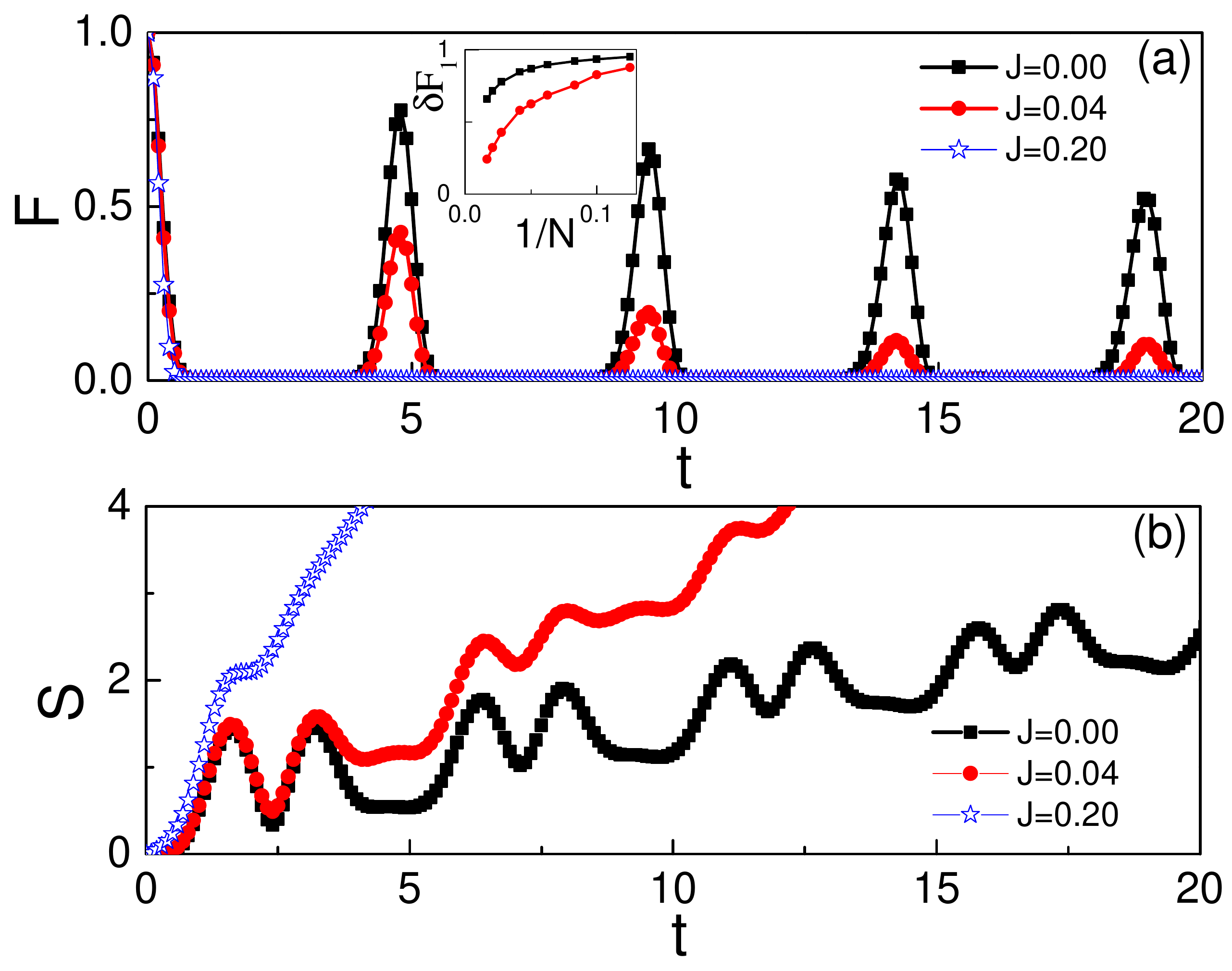}
\caption{Quantum characteristics of the $S=1$ Kitaev-Heisenberg chain
of $N=36$ sites, starting from $\vert -+-+ \cdots -+ \rangle$ in Eq.
\eqref{Ham7} as the initial state for selected values of $J$:
(a) the fidelity $F$ and
\mbox{(b) the entanglement entropy ${\cal S}$.}
The inset in (a) shows the scaling of the first revival
peak $\delta F_1$ versus $1/N$ for $J=0.0$ and $J=0.04$.
The bond dimension is set as $\chi=400$. }
\label{F_S_spin1_KH_N36}
\end{figure}

We note that the phase diagram of the $S=1$ Kitaev-Heisenberg chain
resembles to some extent the phase diagram of the Kitaev-Heisenberg
chain for $S=\frac12$ spin \cite{Got17}, and one finds the following
correspondence between phases:
the Haldane phase for $S=1$ corresponds to the AFM phase for
$S=\frac12$, and the FM phase for $S=1$ corresponds again to the FM
phase when spin is smaller. But it is remarkable that the LLRR phase
corresponds to two distinct phases when $S=\frac12$: the zigzag phase
for $K>0$ and the stripe phase for $K<0$. Thereby the regime of the LLRR
phase for $K>0$ extends over a broader range of angle $\phi$, similar to
the broader range of $\phi$ for the zigzag phase at $S=\frac12$
\cite{Got17}.
In addition, both the FM and the Haldane phase consist here of the two
distinct phases, depending on whether $K>0$ or $K<0$. This demonstrates
that the quantum components of $S=1$ spins are nonequivalent and the
SU(2) symmetry of the Heisenberg term does not hold when Kitaev
interactions are finite.

Finally,
we investigate the scar stability in a broader regime, which is
relevant for possible solid state applications. Then we consider scar
stability of the Kitaev model with Heisenberg perturbations, and show
that they display anomalous stability in the Kitaev phase. Figure
\ref{F_S_spin1_KH_N36} shows the fidelity with periodic slowly-decaying
revivals in the dynamics for $J=0$. In contrast, it completely collapses
for $J=0.2$ with $N=36$. The coherent dynamics is retained in the Kitaev
phase while it is lost quickly when the phase enters other phases. Away
from the Kitaev point, one can observe how the entanglement grows
rapidly with time. These results suggest that exact scars are a generic
property of Kitaev models.

\section{Summary and conclusions}
\label{sec:summa}

In this paper, we studied the eigenstate properties of
Hamiltonian relevant for spin-orbit-coupled electronic system, where the
PXP model was embedded. Thus, the many-body quantum scars and the
associated constrained dynamics can be unveiled. This phenomenology is
characterized by the fact that the dynamics is anomalously slow provided
the initial state has a non-negligible overlap with scarred states. The
ETH violations have been explicitly detected in the uniform $w_j=1$
($\forall j$) sector, where the spin-1 Kitaev model can be rigorously
mapped into a spin-$\frac{1}{2}$ PXP model.  The PXP model acts here as
an archetypal model to possess quantum scars, which are introduced to
explain the slow dynamics observed by evolving a charge-density wave
initial state in the Rydberg atom chain realized in Ref.~\cite{Bernien17}.
As a consequence of nearest-neighbor Rydberg blockaded, the dimension
of constrained Hilbert space grows as $g^N$, where $g=(1+\sqrt{5})/2$
is the golden ratio. A few scarred eigenstates spread throughout the
spectrum, and their presence is evidenced by sub-extensive entanglement
entropy and persistent oscillation of local observables. We have shown
that there is coexistence of volume-law and area-law entangled
eigenstates throughout the spectrum.

These ETH violating eigenstates survive despite hybridization with an
exponential number of thermal eigenstates and motivated the study of
their stability against perturbation. We consider the perturbations by
Heisenberg interactions, see Eq. \eqref{Ham10}. In order to characterize
the ground state properties, we adopt both the local and nonlocal
correlations, which identify distinct phases when the Kitaev coupling
$K$ and the Heisenberg coupling $J$ vary, see Fig. \ref{PD_chart}. For
large negative $J$, the FM order along $z$ ($x$) axis is favored for
$K=1$ ($K=-1$). Increasing the value of $J$, the ground state evolves
from the FM state into the stripy phase, in which the spin structure has
a left-left-right-right pattern. The magnetic order vanishes when $J$
approaches the Kitaev limit. In stark contrast to the gapless ground
state of $S=\frac{1}{2}$ Kitaev chain, the gapped ground state supports
a narrow Kitaev phase in the vicinity of $J=0$, which is characterized
by the extremely short-range correlations.

As a universal feature, the system enters into the Haldane phase upon
increasing $J$. The Haldane phase maintains its topological character
and cannot evolve adiabatically to other phases, since it is protected
by the combination of the spatial inversion symmetry, the time-reversal
symmetry, and the dihedral $D_2$ symmetry. The continuous revivals for
starting from particular initial states remains in the disordered
Kitaev phase, while the state coherence breaks down swiftly in
symmetry-breaking phases. Nevertheless, with the help of examples of
single-ion term and multispin interactions, we demonstrate the
scarred states appear to be robust against $\mathbb{Z}_2$-symmetry
preserving perturbations, at least for currently experimentally
accessible system-sizes. Our findings highlight
that spin-1 Kitaev systems exhibit a rich variety of phenomena
and thus provide a generic non-integrable constrained quantum many-body
system to study non-ergodic behaviors, the stability under generic
perturbations and quantum analogues of the Kolmogorov-Arnold-Moser theorem.

\begin{acknowledgements}
The authors appreciate very insightful discussions with \mbox{Wojciech}
Brzezicki, Hosho Katsura, and Zhi-Xiang Sun. This work is supported by
the National Natural Science Foundation of China (NSFC) under Grants
No. 12174194, the startup fund of Nanjing University of Aeronautics and
Astronautics Grant No. 1008-YAH20006, Top-notch Academic Programs Project
of Jiangsu Higher Education Institutions (TAPP) and stable supports for
basic institute research, Grant No. 190101. \mbox{A. M. Ole\'s}
kindly acknowledges support by Narodowe Centrum Nauki (NCN, Poland)
under Project No. 2016/23/B/ST3/00839 and is grateful for support by
the Alexander von Humboldt Foundation \mbox{(Humboldt-Forschungspreis).}
\end{acknowledgements}

\appendix

\section{Bipartite entropy of two-site spin-1/2 PXP model and spin-1 Kitaev model}
\label{bipartiteentropy_N2}

Here we consider $N=2$. The Hamiltonian of spin-1/2 PXP model within the
bases in the constrained Hilbert space,
\begin{eqnarray}
\label{basisPXPN2}
\{\vert{\downarrow\downarrow}\rangle,
  \vert{\downarrow\uparrow}\rangle,
  \vert{\uparrow\downarrow}\rangle\},
\end{eqnarray}
\begin{eqnarray}
\label{H_PXP_N2}
H=\left(
   \begin{array}{ccc}
     0 & 1 & 1 \\
     1 & 0 & 0 \\
     1 & 0 & 0 \\
   \end{array}
 \right),
\end{eqnarray}
and the ones for the spin-1 Kitaev model within $(w_i=1$,
\mbox{$i=1,2,\cdots,N)$} subspace
\begin{eqnarray}
\label{basisKitaevN2}
\{\vert -+ \rangle, \vert +-  \rangle,  \vert  00 \rangle\} ,
\end{eqnarray}
\begin{eqnarray}
\label{H_Kitaev_N2}
H=\left(
   \begin{array}{ccc}
     0 & 0 & 1 \\
     0 & 0 & 1 \\
     1 & 1 & 0 \\
   \end{array}
 \right).
\end{eqnarray}

There is a one-to-one correspondence between Eqs. (\ref{basisPXPN2})
and (\ref{basisKitaevN2}):
 \begin{eqnarray}
\vert -+\rangle\Rightarrow \vert{\downarrow\uparrow}\rangle,
\vert +-\rangle\Rightarrow\vert{\uparrow\downarrow}\rangle,
\vert 00\rangle\Rightarrow\vert{\downarrow\downarrow}\rangle.
\end{eqnarray}
 The eigenvectors of Eq. (\ref{H_PXP_N2})
 \begin{eqnarray}
\vert\psi_1\rangle&=&-0.7071\,\vert{\downarrow \downarrow}\rangle+0.5000\,
\vert{\downarrow\uparrow}\rangle +0.5000\,\vert{\uparrow\downarrow}\rangle, \nonumber \quad\\
\vert\psi_2\rangle&=&\;\;\,0.0000\,\vert{\downarrow\downarrow}\rangle-0.7071\,
\vert{\downarrow\uparrow} \rangle +0.7071\,\vert{\uparrow\downarrow}\rangle, \nonumber \quad\\
\vert\psi_3\rangle&=&\;\;\,0.7071\,\vert{\downarrow \downarrow}\rangle+0.5000\,
\vert{\downarrow\uparrow} \rangle +0.5000\,\vert{\uparrow\downarrow}\rangle. \quad
\end{eqnarray}

The entanglement spectra and von Neumann entropy
  \begin{eqnarray}
\lambda_{1,3}^{(1,2)}&=&\frac{1}{2}\pm\frac{\sqrt{3}}{4}\,, \quad {\rm with}\quad S_A=0.3546,\\
\lambda_2^{(1,2)}&=&\frac{1}{2},\frac{1}{2}, \quad\quad\quad {\rm with}  \quad S_A=1.0.
\end{eqnarray}
Similarly,  the eigenvectors of Eq. (\ref{H_Kitaev_N2})
  \begin{eqnarray}
\vert\phi_1\rangle
 &=&0.5000\,\vert{-+}\rangle+0.5000\,\vert{+-}\rangle -0.7071\,\vert 00\rangle, \nonumber \\
\vert\phi_2\rangle
 &=&0.7071\,\vert{-+}\rangle-0.7071\,\vert{+-}\rangle +0.0000\,\vert 00\rangle, \nonumber  \\
\vert\phi_3\rangle
 &=&0.5000\,\vert{-+}\rangle+0.5000\,\vert{+-}\rangle +0.7071\,\vert 00\rangle.
\end{eqnarray}
The entanglement spectra and von Neumann entropy are:
\begin{eqnarray}
\lambda_{1,3}^{(1,2,3)}&=&  \frac{1}{4},\frac{1}{4},\frac{1}{2} , \quad {\rm with} \quad  S_A=1.5,\\
\lambda_2^{(1,2,3)}&=&\frac{1}{2}, \frac{1}{2}, 0,\, \quad {\rm with} \quad  S_A=1.0.
\end{eqnarray}
To this end, the diverse choice of bases in Eqs. (\ref{basisPXPN2}) and
(\ref{basisKitaevN2}) will lead to multiple values of bipartite
entanglement entropy.

\begin{figure}[t!]
\includegraphics[width=\columnwidth]{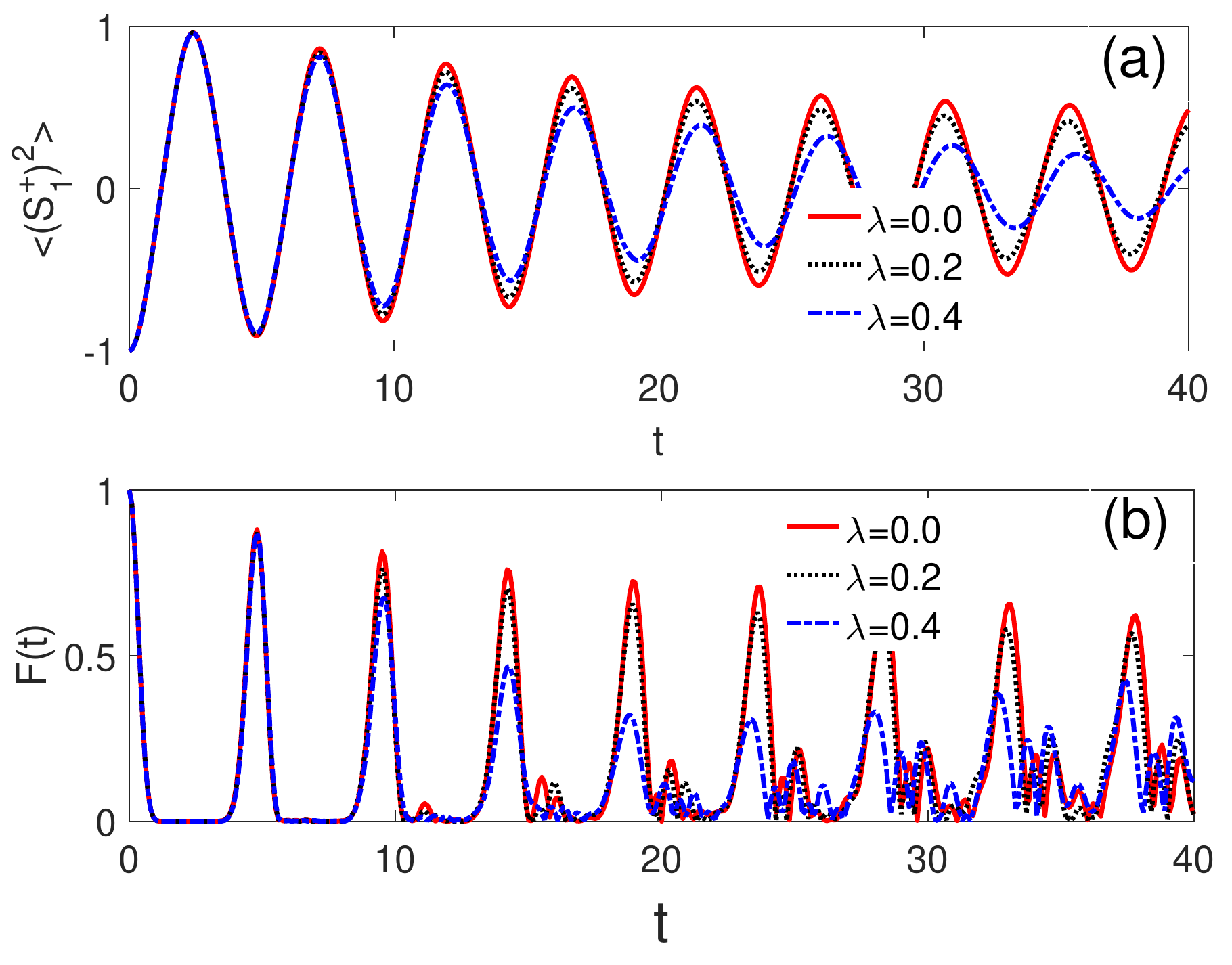}
\caption{Quantum properties of the Kitaev chain with $N=18$ sites
starting from the initial $|\tilde{\mathbb{Z}}_2\rangle$ state with
XZY interactions and for $\lambda=0.0$, 0.2, 0.4:
(a) the order parameter \eqref{O1}, and (b) the quench fidelity $F(t)$.}
\label{F_OP_t_XZY}
\end{figure}

\section{The Kitaev chain under XZY-type perturbations}
 \label{Kitaev_XZY}

The energy current follows from the continuity equation~\cite{Zotos1997},
\begin{equation}
\hat{J}_{\rm E} = \sum_{j=1}^{N} i [h_j,h_{j+1}] = K^2
\sum_{j=1}^{N} S_{j}^x S_{j+1}^z S_{j+2}^y \, .
\label{Eq2}
\end{equation}
This operator acts on three adjacent sites and contains also the
$z$ component of spin-$1$ operators, which commutes with $\tilde{W}_j$.
The presence of an effective energy flow
will manifest itself in the effective Hamiltonian followed
by a Lagrange multiplier~$\lambda$:
\begin{eqnarray}
\tilde{H}_{\rm K}=K \sum_{j=1}^{N}~S^x_jS^y_{j+1}-\lambda \hat{J}_{\rm E}.
 \label{ham4}
\end{eqnarray}
It can be verified that
 $[\tilde{W}_j, \tilde{W}_k]=0$,   %{\rm and \quad}
 $[\tilde{W}_j, \tilde{H}_{\rm K} ]=0$.
In this respect, the three-site XZY-type interactions belong to
\mbox{$\mathbb{Z}_2$-symmetry} preserved perturbations.
The numerical results for different values of $\lambda$ are shown in
Fig. \ref{F_OP_t_XZY}.
One finds that the oscilations of the order parameter \eqref{O1} are
more damped and the fidelity $F(t)$ has a richer structure at larger
time $t$ when $\lambda$ increases.

\end{document}